\documentclass[useAMS,usenatbib]{mn2e}
\usepackage[dvips]{graphicx}

\title[A more detailed look at the Opacities for Enriched Carbon and Oxygen Mixtures.]{A more detailed look at the Opacities for Enriched Carbon and Oxygen Mixtures.}

\author[J. J. Eldridge and C. A. Tout] {J.J. Eldridge\thanks{E-mail: jje@ast.cam.ac.uk} and C.A. Tout \\ Institute of Astronomy, Madingly Road, Cambridge, CB3 0HA, England\\}

\date{Written 1/9/2003}
\pagerange{\pageref{firstpage}--\pageref{lastpage}} \pubyear{2003}
\begin{document}
\maketitle
\label{firstpage}

\begin{abstract}
We have included opacity tables in our stellar evolution code that enable us to accurately model the structure of stars composed of mixtures with carbon and oxygen independently enhanced relative to solar. We present tests to demonstrate the effects of the new tables. Two of these are practical examples, the effect on the evolution of a thermally pulsing asymptotic giant branch star and a Wolf-Rayet Star. The changes are small but perceptible.
\end{abstract}


\section{Introduction.}
The main source of a star's energy is the nuclear fusion reactions occurring either in its core or in thin burning shells around the core. This energy is transported from the production site to the surface by radiative transfer or convection. The first transports energy in the form of photons, while the second is a cyclic macroscopic mass motion that carries the energy in bulk.

In regions where convection is stabilised, radiative transfer leads to the equation of stellar structure,

\begin{equation}
\frac{dT}{dr} = - \frac{3 \overline{\kappa_{\rm R}} \rho}{16 \sigma T^{3}} \frac{L_{r}}{4 \pi r^{2}},
\end{equation}

where $L_{r}$ is the luminosity at radius $r$, $\rho$ the density, $\sigma$ the Stefan-Boltzmann constant, $T$ the temperature and $\overline{\kappa_{{\rm R}}}$ is the Rosseland mean opacity.

Opacity $\kappa$ is a measure of the degree to which matter absorbs photons. There are four main sources of opacity in stars

\begin{itemize}
\item Bound-bound transitions are the transitions of an electron in an atom, ion or molecule between energy levels which are accompanied by either the absorption or emission of a photon. Only a photon with the correct wavelength can cause a given transition, the process is wavelength dependent.
\item Bound-free transitions, or photoionisation, occur when an incoming photon has enough energy to ionise an atom or ion and free an electron. The reverse process is the capture of an electron by an atom or ion. It will not occur until photons above a threshold energy are available and falls off as $\nu^{-3}$ where $\nu$ is the frequency.
\item Free-free transitions are scattering processes which occur when an electron and photon interact near an atom or ion. The process is also known as bremsstrahlung. Again it is proportional to $\nu^{-3}$.
\item Electron scattering is wavelength independent at low temperatures where it is Thomson scattering. The electron, with its low cross-section, is a small target and so only dominates at high temperatures when most atoms are ionised. At very high temperatures relativistic effects are important and Compton scattering dominates.
\end{itemize}

From this list it is possible to see that calculating the opacity in a stellar model is a difficult process and depends on the composition, temperature and density of the material. Deep in the star local thermodynamic equilibrium (LTE) is achieved and an average for radiative transfer over all wavelengths requires the Rosseland mean opacity $\overline{\kappa_{{\rm R}}}$ \citep{RO24} expressed as

\begin{equation}
\frac{1}{\overline{\kappa_{{\rm R}}}} =\frac{ \displaystyle \int_{0}^{\infty} \frac{1}{\kappa_{\nu}} \frac{\partial B_{\nu}}{\partial T} d\nu }{\displaystyle \int_{0}^{\infty} \frac{\partial B_{\nu}}{\partial T} d\nu},
\end{equation}

where $\kappa_{\nu}$ is the opacity at frequency $\nu$, $T$ is the temperature and $B_{\nu}$ is the flux per unit area into unit solid angle per unit frequency in LTE.

The most comprehensive opacities available today are from the OPAL \citep{IR96} or OP \citep{OP} groups who have made detailed  models of the above processes. They provide tables of the Rosseland mean opacity variation with temperature, density and composition with the metal abundance usually scaled to solar compositions. \citet{IR93} took a step forward for the OPAL project team by providing tables that include mixtures enhanced in carbon and oxygen (C~and~O) relative to the base solar composition. We have incorporated their full range of composition tables in the \citet{E71} evolution code. In the last implementation of this code \citep{P95} there only 10 of the 265 tables available were used. This provides us with the opportunity to refine our models so as to accurately follow the changes of opacity which occur in the later stages of evolution. The effects are expected in many types of stars. While these may be small for the evolution of main-sequence stars and white dwarfs we expect larger differences to be found in AGB and Wolf-Rayet stars.

\begin{itemize}
\item Asymptotic Giant Branch (AGB) stars undergo third dredge-up which mixes helium burning products to the surface and forms carbon stars. Using the enhanced mixture tables we shall be able to model the thermal pulses and envelope evolution more accurately.
\item Wolf-Rayet (WR) stars are massive and have lost their hydrogen envelopes exposing the helium cores. As time progresses helium burning products are slowly exposed at the surface and in some cases the stars are eventually mostly composed of carbon and oxygen.
\end{itemize}

We present our method of opacity table construction and detail its implementation in the Eggleton stellar evolution code. We discuss the effects on main-sequence stars, red giants and white dwarfs. We then present three tests of our work. The first is the effect of including extra carbon and oxygen on the structure and evolution of a low-mass population-III star, the second is a $5{\rm M}_{\odot}$ thermally pulsing AGB star and the third a Wolf-Rayet star of $40{\rm M}_{\odot}$  with mass loss.

\section{Table Construction.}

We start with the OPAL tables \citep{IR96} as the framework around which we construct our full tables in an approach similar to that employed by \citet{P95}. However we choose to use the variable ${\mathcal{R}}=\rho / T^{3}_{6}$ rather than the density. This has the advantage that our tables can be a third smaller in this dimension. This partly compensates for increased memory requirement of the full tables.

The OPAL tables extend in $\log_{10} (T/{\rm K})$ from $3.75$ to $8.70$ in steps of $0.05$ to $0.2$ and in $\log_{10} ({\mathcal{R}}/ {\rm g \, cm^{-3} K^{-3}})$ from $-7$ to $+1$ in steps of $0.5$. We choose the range of the final tables to cover from $3$ to $10$ in $\log_{10} (T/{\rm K})$ with increments of $0.05$, and from $-8$ to $+7$ in $\log_{10} ({\mathcal{R}}/ {\rm g \, cm^{-3} K^{-3}})$ with increments of $0.5$. This gives us tables of 141 by 31 elements.

To account for the effect of composition, the OPAL tables include typically 265 $T$~and~$\mathcal{R}$ grids for each metallicity. These tables are split into 5 groups with different hydrogen mass fractions X of $0.0$, $0.03$, $0.1$, $0.35$ and $0.7$. The tables in each group have different compositions in helium, carbon and oxygen.

The lowest temperature in the OPAL tables is $\log_{10} (T/{\rm K}) = 3.75$ and we use the tables of \citet{AF94}, which extend from $3$ to $4.1$ in $\log_{10} (T/{\rm K})$, to complete them at low temperature. Where the tables overlap they match well \citep{IR96}. It should be noted that the tables provided by \citet{AF94} do not include enhanced carbon and oxygen mixtures. Therefore at temperatures when $\log_{10} (T/{\rm K}) < 4.0$ our tables do not follow these mixtures. This is only important in the surface of AGB stars. We plan in future to deal with low temperature enhanced mixtures by including the work of \citet{M02} before application to the structure of AGB star envelopes.

To complete each table we fill the region from $8.70$ to $10$ in $\log_{10} (T/{\rm K})$, where electron scattering dominates, according to \citet{BY76}. We then combine with a full table of the effective opacity owing to electron conduction for which we use the fits of \citet{I75} to the tables of \citet{HL69} and \citet{Ca70}. We combine the radiative opacity with the effective conductive opacity by reciprocal addition,

\begin{equation}
\frac{1}{\kappa_{\rm eff}}=\frac{1}{\kappa_{\rm rad}}+\frac{1}{\kappa_{\rm cond}}.
\end{equation}

Before the complete tables are implemented in the stellar evolution code we perform one last process. We rescale the tables over the C and O plane to a more regular grid. In the OPAL grid the spacing used requires interpolation within irregular polygons, so to make the interpolation simpler and faster, we include extra tables in a similar but fully rectangular grid system. This means we have 61 tables for each X abundance or 305 tables for each metallicity.

\section{Implementation into a Stellar Evolution Code.}

We implement the tables in the stellar evolution code first described by \citet{E71} and most recently updated by \citet{P95}. Further details can be found in the references therein. The main drawback with implementation of the new tables is the increased memory requirement. Whereas before there were only 10 opacity tables there are now 305. The new tables are each a third of the size of the old tables so the overall increase is a factor of ten. This corresponds to a memory requirement increase from 17$\,$Mb to 170$\,$Mb but the average specification of today's desktop computers comfortably accommodates this. One of the reasons why this memory requirement is so large is that we also store the spline coefficients for each of the tables. When we use the tables within the code we interpolate in the $\mathcal{R}$ and $T$ plane with bicubic splines. This ensures the smoothness of our tables. We then use linear interpolation in the 3 composition dimensions to derive the final opacity value at each point.

To test the effect of these tables we use three different interpolation schemes,

\begin{itemize}
\item Method A uses 5 tables with $X=0,\,0.03,\,0.1,\,0.35$ and $0.7$ and $Y=1-X-Z$. Where $X$ is the hydrogen mass fraction, $Y$ the helium mass fraction and $Z$ the initial, solar mixture, metallicity. We add a further two tables, both with $X=Y=0$, the first with $X_{\rm C}=1-Z$ and $X_{\rm O}=0$. The other with $X_{\rm C}=0$ and $X_{\rm O}=1-Z$. Where $X_{\rm C}$ and $X_{\rm O}$ are the mass fractions for the enhanced amount of carbon and oxygen above that included in the metallicity mass fraction so the total carbon mass fraction is $X_{\rm C} + Z({\rm C})$ and for oxygen $X_{\rm O} + Z({\rm O})$. This gives a total of 7 tables allowing simple interpolation in composition similar to the old method.
\item Method B is similar to method A but we include two extra tables for each hydrogen mass fraction. Again both have $Y=0$. One has $X_{\rm C}=1-X-Z$ and $X_{\rm O}=0$ and the other $X_{\rm C}=0$ and $X_{\rm O}=1-X-Z$. This leads to a total of 15 tables in all.
\item Method C incorporates all 305 tables in the opacity calculations. We have 61 tables for each hydrogen abundance. When interpolating in composition we use three variables $X$, $X_{\rm C}/(1-X-Z)$ and $X_{\rm O}/(1-X-Z)$. The hydrogen abundances are as for method A. We take values of $0,\,0.01,\,0.03,\,0.1,\,0.2,\,0.4,\,0.6$ and $1$ for the carbon and oxygen grid planes. On this grid we perform a three-dimensional linear interpolation.
\end{itemize}

\section{Testing \& Results.}

The new tables give almost identical results to the old for main-sequence stars and red giants. The timescale of core helium burning is altered. From method A to method C it increases by a factor of $0.1$ percent for a $5{\rm M}_{\odot}$ star. This is reassuring. Current stellar evolution models describe well the evolution of the main sequence and the red giant branch. However we do find that the opacity  tables enhance numerical stability in the later stages of evolution because they resolve changes with composition in greater detail. This leads to a smoother variation in opacity.

A similar result is found for white dwarfs. Between methods~A~and~C for a carbon and oxygen white dwarf the radius increases by 0.2 percent. While conditions in the atmosphere do pass through regions of the greatest difference between the two methods (the dotted line in figure \ref{fc}) there is not enough mass at these temperatures and densities to make a significant global difference.

\subsection{Polluting a $0.5{\rm M}_{\odot}$ Population-III Star.}

For our first detailed test we take a zero metallicity $0.5 \, {\rm M}_{\odot}$ star and pollute it uniformly with carbon and oxygen. Such a star could have been formed just after the death of the first stars if the products of these stars were mainly carbon and oxygen. We choose such a low-mass star to limit the effect on the results of burning via the CNO cycle. Our initial model has the following properties, $X=0.7$, $Y=0.3$, $R=0.45 \, {\rm R}_{\odot}$, $T_{{\rm eff}}=4780\,{\rm K}$ and $L=0.96 \, {\rm L}_{\odot}$. The tables provided by OPAL do not include one with $X=0.75$, we could extrapolate to obtain the opacity but for this test we prefer to remain within the tables. From the initial model we perform three sub-tests. In the first we pollute the star at a constant C/O ratio of 1, then evolve each model and compare the time from zero-age main sequence to the helium flash for methods A, B and C. We also compare a sequence of models for stars of the same mass but with a metallicity of the same mass fraction as the carbon and oxygen. These results are presented in table \ref{ta}. We see that with the pollution fraction increase, the timescale of hydrogen burning increases because the energy produced at the centre takes longer to reach the surface. However when comparing the difference between methods B \& C we see only a comparatively small increase in timescale. When we compare with the scaled solar models we see a greater difference. This is due to the much larger opacity of heavier elements such as iron.

With the second set of models we look at the radius, surface temperature and luminosity of the star on the main sequence as we pollute it with either only carbon or only oxygen (tables \ref{tb} and \ref{tc}). We use method A as the base to which we compare methods B and C. A similar trend is seen in the results in tables \ref{tb} and \ref{tc}. We therefore see it is important to include the affect of enhanced C and O mixtures before hydrogen exhaustion. Similar mixtures are encountered in the envelopes of carbon stars of very low metallicity.

Finally we consider the effect of varying the C/O ratio of the pollution (table \ref{td}) at a constant pollutant mass fraction of $0.001$. Table \ref{td} demonstrates the need to use method C to follow accurately the variation in opacity as the C/O ratio changes. This is important in AGB evolution during the thermal pulses when the C/O ratio changes gradually as carbon becomes dominant. The difference between methods B and C in this case comes about because of non-linear structure in the C and O opacities that can only be resolved with the full set of tables.

\subsection{Thermal Pulses of an AGB Star.}

Our thermally pulsing AGB model is of a $5{\rm M}_{\odot}$ star. To follow a number of pulses we include convective overshooting \citep{b10} during the pre-AGB evolution. This leaves the star with a larger core which reduces the rate at which the thermal pulses grow in strength. This means we do not find dredge-up in our models but we can evolve through the pulses. We obtain the average pulse period as the arithmetic mean of the times between the first ten helium shell flashes. Results are recorded in table \ref{te}.

First we find that the different methods affect the age of the star when it starts to undergo thermal pulses. Second we find that methods A and B have similar interpulse periods while method C's is $3.6$ percent less. We attribute this to the change in opacity due to varying He/C/O ratios in the inter-shell region which alters the time evolution of the pulse.

\subsection{Evolution of a Wolf-Rayet Star. }

We have used the mass-loss rates of \citet{NL00} to manufacture a Wolf-Rayet star of solar metallicity and an initial mass of $40{\rm M}_{\odot}$. Some details are described by \citet{DT03}. Our tables are expected to make a difference because Wolf-Rayet stars become naked helium stars and helium burning products are mixed throughout and exposed at the surface of the star. We only use methods A and C and not method B because the carbon and oxygen mass fractions do not rise above those included in the base metallicity until after the hydrogen envelope has been removed and so our tables only begin to make a difference after this.

\begin{figure}
\includegraphics[height=84mm,angle=270]{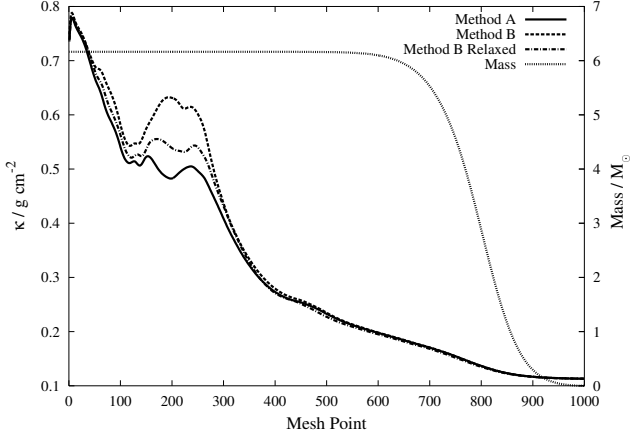}
\caption{The variation of opacity throughout the Wolf-Rayet star just after core carbon ignition. The solid line is calculation from model A, the dashed line from model C and the dash-dotted lines from model C with hydrostatic equilibrium. The dotted line is the mass interior to the mesh point.}
\label{fa}
\end{figure}

\begin{figure*}
\includegraphics[height=170mm]{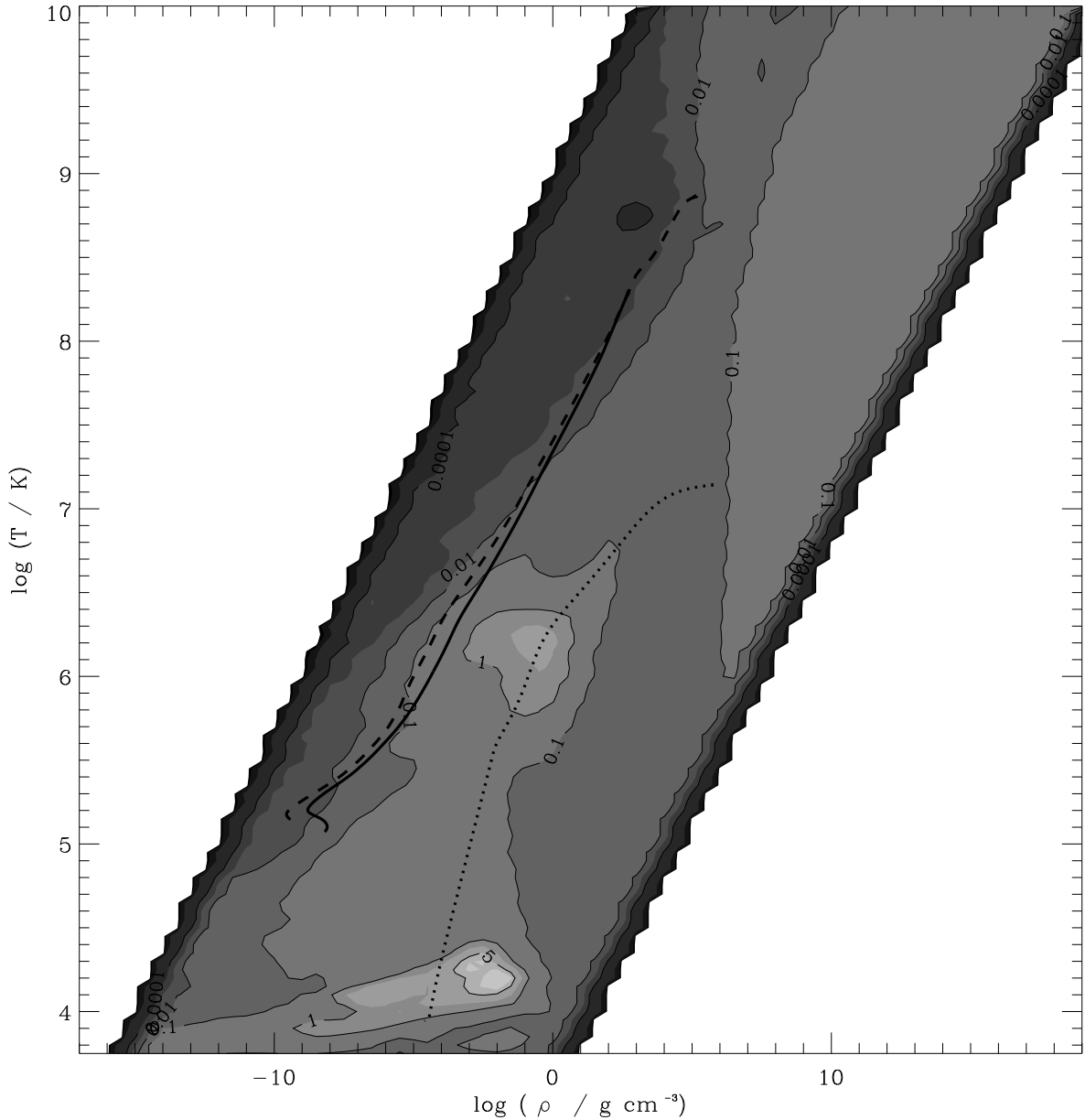}
\caption{The value plotted by the contours is $ \int_{S} (\kappa_{\rm C}-\kappa_{\rm A}) \, dX_{\rm C} \, dX_{\rm O} / \int_{S} \kappa_{0} \, dX_{\rm C} \, dX_{\rm O}$. Here $\kappa_{\rm A}$ and $\kappa_{\rm C}$ are the opacities calculated by methods A and C. The opacities are all calculated when $X=0$ and over the surface ($S$) defined by $(X_{\rm C}+X_{\rm O})=0$ to $1$. The vertical artifacts are from the wide spacing of the tables in $\rho$. We show two Wolf-Rayet models on this plot, solid line $12{\rm M}_{\odot}$ and dotted $6.1{\rm M}_{\odot}$, in the same stage of evolution as in figure \ref{fa}. We see that the largest differences in the models indeed occur as the star drifts into the region where the opacity difference is greatest. Also plotted is the dotted line for a $0.3{\rm M}_{\odot}$ pure C/O white dwarf model. While the line passes through regions of the diagram with greater values the white dwarf is only slightly affected because very little mass is in the regions with greatest difference.}
\label{fc}
\end{figure*}

In figure \ref{fa} we present the variation of opacity through the star, just after core carbon ignition, with mesh point, the grid over which we solve the equations of stellar structure. The solid line is with method A. We then take this structure and use method C to calculate the opacities to give the dashed line. This model is not in hydrostatic equilibrium and the dot-dashed line shows the relaxed new structure in hydrostatic equilibrium from method~C. Comparing the physical parameters of these calculations we find the main change to be an increase in radius of 3.0 percent. The surface temperature has decreased by 1.5 percent.

\begin{figure}
\includegraphics[height=84mm,angle=270]{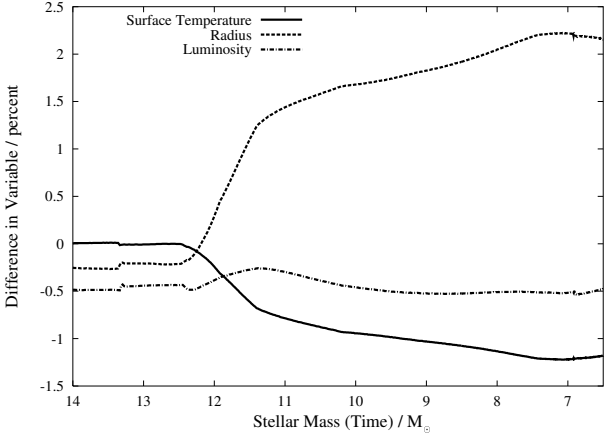}
\caption{The percentage difference in model parameters between methods A and C versus mass of the Wolf-Rayet star.}
\label{fb}
\end{figure}

We then made two complete runs with similar initial conditions by methods~A~and~C. In figure \ref{fb} we display the difference between Wolf-Rayet evolution as the hydrogen envelope is removed. The greatest changes are in the radius and surface temperature. The star is in the WN phase between $14$ to $12 {\rm M}_{\odot}$, after which it enters the WC phase. In the WN phase there are only small differences but upon entering the WC phase the differences build up so that the temperature is lower by around 2 percent and the radius about 1 percent larger. The luminosity also tends to be larger by around 0.5~~percent.
Figure \ref{fc} is a plot of where the difference between methods A and C are greatest in $\rho$ and $T$ when $X=0$. The plot is of the mean difference in the opacity calculated by methods A and C at all possible carbon and oxygen abundances divided by the opacity when $Y=1-Z$. From the plot we see that the regions of greatest difference are small. The lines show where the Wolf-Rayet model calculated by method C lies at certain points of its evolution. The solid line is for the star at $12.3{\rm M}_{\odot}$ and the dashes at $6.1{\rm M}_{\odot}$. From figure \ref{fb} we see that the largest changes begin around the $12{\rm M}_{\odot}$ point. This is when the model has moved into the region where the differences between A and C are greatest.

We find the opacity tables alter the timescales for the late burning stages. Method C increases the helium burning timescale by 0.1 percent while the carbon-burning time scale increases by 2.5 percent relative to method A. These changes come about because the opacity varies with composition and effects the temperature structure and burning rates. While there is no large difference in a single variable the overall effect on structure and evolution is more significant.

\section{Conclusions.}

Our major conclusion is that the changes induced by properly including opacities for varying C and O mixtures are small. However they are a good thing to include in stellar evolution codes because they help numerical stability by removing sharp changes in the variation of opacity with composition that are encountered when we use fewer tables and add another level of detail to models without much loss of computational speed.

The main computational cost is the extra memory to store the tables. Method C requires ten times the memory (about 200$\,$Mb) compared to methods A and B. However there is little computational speed cost: relative to method A, methods B and C require about 4 percent more time for the evolution of a Wolf-Rayet star even though we must evaluate four times as many opacities for each $\mathcal{R}$ and $T$ value than with method A.

The tables and stellar evolution code are freely available from http://www.ast.cam.ac.uk/$\sim$stars. Interested users are welcome to contact the authors for details on how to download and implement our tables.

\section{Acknowledgements.}

The authors would like to thank Onno Pols for informative suggestions. JJE would like to thank PPARC for his funding and his parents for helping to provide the computing power for this work. CAT thanks Churchill College for his fellowship.

\label{lastpage}
\bsp
\newpage

\begin{table}
\caption{Variation of Evolutionary Time Scale with increasing carbon and oxygen mass fraction. Here $t_{\rm A}, t_{\rm B}$ and $t_{\rm C}$ are the evolution times to the helium flash for methods A, B and C and $t_{\rm Z}$ is the evolution time for models with an equivalent metallicity.}
\label{ta}
\begin{tabular}{ccccc}
$(X_{\rm C}+X_{\rm 0})$ & $\frac{t_{\rm C}-t_{\rm A}}{t_{\rm A}}$ & $\frac{t_{\rm B}-t_{\rm A}}{t_{\rm A}}$ &  $\frac{t_{\rm B}-t_{\rm C}}{t_{\rm A}}$&  $\frac{t_{\rm Z}-t_{\rm C}}{t_{\rm C}}$\\
or $Z$ Mass  & /$10^{-2}$ & /$10^{-2}$ & /$10^{-2}$  & /$10^{-2}$\\
Fraction & & & & \\
\hline
$10^{-5}$ & 0.035 & 0.031 & 0.004 & 0.124\\
$10^{-4}$ & 0.355 & 0.311 & 0.045 & 1.134\\
$10^{-3}$ & 3.535 & 3.047 & 0.488 & 10.02\\
$10^{-2}$ & 32.94 & 31.49 & 1.449 & 57.54\\
\end{tabular}
\end{table}

\begin{table}
\caption{Variation in physical parameters relative to method A with increasing carbon mass fraction.}
\label{tb}
\begin{tabular}{ccccc}
 & & $\Delta R / R$ & $\Delta T_{\rm eff} /T_{\rm eff} $ & $\Delta L / L$ \\
 & $X_{\rm C}$ & $/10^{-2}$ & $/10^{-2}$ & $/10^{-2}$\\
\hline
(B-A)/A & $10^{-6}$ & 0.000 & 0.000 & 0.005\\				
 & $10^{-4}$ & 0.046 & 0.071 & 0.384 \\	
 & $10^{-3}$ & 0.485 & 0.654 & 3.533 \\	
 &$10^{-2}$ & 4.124 & 3.439 & 21.094 \\
(C-A)/A & $10^{-6}$ & 0.000 & 0.000 & 0.002 \\				
 & $10^{-4}$ & 0.041 & 0.060 & 0.326 \\	
 & $10^{-3}$ & 0.433 & 0.563 & 3.058 \\	
 & $10^{-2}$ & 4.168 & 3.457 & 21.219 \\
(B-C)/A & $10^{-6}$ & 0.0000 & 0.0000 & 0.0023 \\				
 & $10^{-4}$ & 0.0046 & 0.0115 & 0.0574 \\	
 & $10^{-3}$ & 0.0619 & 0.0915 & 0.4743 \\	
 & $10^{-2}$ & -0.0441 & -0.0178 & -0.1250\\
\end{tabular}
\end{table}

\begin{table}
\caption{Variation in physical parameters relative to method A with increasing oxygen mass fraction.}
\label{tc}
\begin{tabular}{ccccc}
 & & $\Delta R / R$ & $\Delta T_{\rm eff} /T_{\rm eff} $ & $\Delta L / L$ \\
 & $X_{\rm O}$ & $/10^{-2}$ & $/10^{-2}$ & $/10^{-2}$\\
\hline
(B-A)/A & $10^{-6}$ & 0.000 & 0.000 & 0.000\\		
 & $10^{-4}$ & 0.009 & 0.023 & 0.113\\		
 & $10^{-3}$ & 0.108 & 0.209 & 1.054\\		
 & $10^{-2}$ & 1.140 & 1.692 & 8.721\\		
(C-A)/A & $10^{-6}$& 0.000 & 0.000 & 0.000\\		
 & $10^{-4}$ & 0.009 & 0.021 & 0.108\\		
 & $10^{-3}$ & 0.106 & 0.200 & 1.013\\		 
 & $10^{-2}$ & 1.179 & 1.710 & 8.859\\		
(B-C)/A & $10^{-6}$ &0.0000&0.0000&0.0000\\		
 & $10^{-4}$ &0.0000&0.0023&0.0046\\		
 & $10^{-3}$ &0.0023&0.0092&0.0410\\		
 & $10^{-2}$ &-0.0387&-0.0181&-0.1386\\		
\end{tabular}
\end{table}

\begin{table}
\caption{Variation in physical parameters relative to method A with varying the C/O ratio.}
\label{td}
\begin{tabular}{ccccc}
 & C/O & $\Delta R / R$ & $\Delta T_{\rm eff} /T_{\rm eff} $ & $\Delta L / L$ \\
 & Ratio & $/10^{-2}$ & $/10^{-2}$ & $/10^{-2}$\\
\hline
B-A & $\frac{1}{3}$ & 2.022 & 2.308 & 12.564 \\		
 & $\frac{1}{2}$ & 2.281 & 2.467 & 13.593 \\	
 & 1     & 2.806 & 2.770 & 15.375 \\			
 & 2     & 3.288 & 3.027 & 17.297 \\
 & 3     & 3.506 & 3.141 & 18.054 \\		

C-A & $\frac{1}{3}$ & 1.891 & 2.200 & 11.948 \\		
 & $\frac{1}{2}$ & 2.121 & 2.348 & 12.886 \\	
 & 1     & 2.685 & 2.687 & 13.080 \\			
 & 2     & 3.223 & 2.985 & 17.040 \\
 & 3     & 3.477 & 3.116 & 17.921 \\		

B-C & $\frac{1}{3}$ & 0.1309 & 0.1080 & 0.6162 \\		
 & $\frac{1}{2}$ & 0.1599 & 0.1191 & 0.7072 \\	
 & 1     & 0.1209 & 0.0829 & 0.4952 \\			
 & 2     & 0.0646 & 0.0424 & 0.2575 \\
 & 3     & 0.0289 & 0.0245 & 0.1322 \\		
\end{tabular}
\end{table}

\begin{table}
\caption{Variation of timescales for a $5{\rm M}_{\odot}$ star.}
\label{te}
\begin{tabular}{ccc}
 & Age at 1st & Average Pulse \\
Method  & Pulse / $10^{8} {\rm yrs}$ & Period / ${\rm yrs}$ \\
\hline
A & 1.199679 & $2{\,}803$ \\
B & 1.193637 & $2{\,}802$ \\
C & 1.200724 & $2{\,}702$ \\
\end{tabular}
\end{table}

\newpage

\end{document}